\pdfoutput=1
\documentclass[%
 aip,
 sd,%
 amsmath,amssymb,
preprint,%
]{revtex4-1}
\usepackage{epsfig}
\usepackage{epstopdf}
\usepackage{graphics}
\usepackage{dcolumn}
\usepackage{bm}
\usepackage{graphicx}
\usepackage{subfigure}

\begin{document}
\title{Effects of blockage, arrangement, and channel dynamics \\ on  performance of turbines in a tidal array}

\author{Xianliang Gong}
\author{Ye Li}%
 \email{ye.li@sjtu.edu.cn.}
 \author{Zhiliang Lin}
\affiliation{ 
School of Naval Architecture, Ocean \& Civil Engineering, Shanghai Jiao Tong University, Shanghai 200240, China
}%

\date{\today}

\begin{abstract}

The performance and economics of turbines in a tidal array are largely dependent on the power per turbine, and so approaches that can increase this power are crucial for the development of tidal energy. 
In this paper, we combine a two-scale partial array model and a one-dimensional channel model to investigate the effects of blockage, turbine arrangement, and channel dynamics on tidal turbines. The power per turbine is obtained as the product of two parameters: a power coefficient measuring the power acquired from the instantaneous flow and an environment coefficient showing the response of the channel to added drag. The results suggest that taking account of channel dynamics will  decrease the predicted power and the optimal induction factor. The model also shows that when the number of turbines in a row is increased, the power per turbine may monotonically increase or decrease or attain a maximum value at a certain global blockage. These different results depend on two characteristic parameters of the channel: $\alpha$ and $\lambda_D$. Furthermore, we find that besides turbine density (blockage), the arrangement of the turbines should also be considered if we want to obtain an efficient array. Appropriate arrangements can enhance the performance of turbines in tidal channels, although the beneficial effects will be partly offset by reduced velocities. The turbine arrangement also has an effect on the optimal global blockage at which the power attains its maximum value. As the rate of increase  of the power per turbine from an array spanning the whole channel width to the optimal partial array diminishes with increasing blockage, the optimal global blockage will also decrease. 
\end{abstract}

\maketitle

\section{Introduction}

Tidal channels are appropriate locations to exploit tidal energy. Positioning turbine arrays in tidal channels can make a huge contribution to the demand for renewable energy. For an array in a channel, there are several aspects of power relating to different perspectives with regard to array performance. \citep{Vennell2012} The power lost to drag, the power lost to the array, and the power available for the array are all measures of array performance that are extremely important in predicting the power potential of a chosen site. However, the decision  whether to build a particular tidal array depends very much on the performance and economics of each turbine in the array. Thus, in this study, we will  focus mainly on the power per turbine.  

Recently, there have been a number of theoretical studies of tidal energy problems. In contrast to  numerical methods, which usually deal with a single aspect, such as turbine blades or wakes, theoretical models can reveal  complex dynamics across a range of different scales from turbine to array and even the whole channel, although at the price of sacrificing some details. For array-scale problems, the focus is generally on the influences of blockage and the arrangement of turbines. For simplicity, a constant flow has normally been assumed in such studies,\citep{GARRETT2007,Houlsby2008,WHELAN2009,Nishino2012a,Nishino2013,Draper2014,Cooke2016} which have made great progress in guiding the design of tidal arrays. \citet{GARRETT2007} (hereinafter GC07) investigated the effects of blockage on power coefficients and turbine efficiency for arrays spanning the whole channel width. They showed that blockage effects  increase the maximum power coefficient by $(1-B)^{-2}$, where $B$ is the blockage ratio. \citet{Houlsby2008} and \citet{WHELAN2009}  considered  free-surface effects, assuming static-pressure conditions. The effects of blockage have  been verified using the RANS model.\citep{Nishino2012} 

Besides blockage effects, turbine performance  can also benefit greatly from an appropriate arrangement of the turbines. \citet{Nishino2012a} (hereinafter NW12) derived a two-scale coupled model to measure the performance of partial arrays by introducing the concept of scale separation between the flow around each device and the flow around the array. Their results showed that there is an optimal inter-turbine spacing from the perspectives of both the power coefficient and efficiency. 
Numerical simulations \citep{Nishino2012} and laboratory experiments \citep{Cooke2015} have given similar results. \citet{Cooke2016} used a three-scale model to describe a partial array  containing many smaller arrays. The above work has mainly paid attention to lateral configurations. \citet{Gong2017} modeled a two-dimensional array that partially block the channel in both the lateral and vertical directions, showing further benefits of an appropriate arrangement. \citet{Draper2014} modeled the dynamics of two-row arrays in the streamwise direction and concluded that arranging turbines in one row is preferable to arranging them in two rows. 

In a wind farm, the atmospheric system is effectively infinite compared with the turbine sweep areas, and thus is hardly affected by the turbines. However, a tidal channel system is more complex and the channel dynamics will be influenced by the turbine operations. According to \citet{Vennell2015},  for most channels, a blockage of $2$--$5\%$ for single-row arrays can influence channel dynamics and this value is smaller for multi-row arrays. Thus, to predict turbine performance, channel-scale analyses are necessary. One appropriate approach to model channel dynamics is to use the one-dimensional shallow-water equation. \citet{Garrett2005} showed that the drag caused by turbines limits the energy potential of channels. \citet{Vennell2010,Vennell2011} then combined the GC07 model with the channel dynamical model to determine the optimal tuning of turbines in whole-width arrays.  

Previous studies of channel dynamics  have emphasized  blockage effects and have used a relatively simple array-scale model (with a whole-width array).\citep{Vennell2010,Vennell2011,Vennell2012,Vennell2013,Vennell2014,Vennell2016} At the same time, those studies that have considered more complex interactions between turbines have usually neglected  interactions between arrays and channels.\citep{Nishino2012,Nishino2013,Draper2014} Thus, when dynamical interactions between arrays and channels are considered, the questions arise as to 
whether there is still an optimal arrangement and, if so, to what extent  the array can benefit from it. 

In this paper, a model will be developed based on previous work to interpret the effects of blockage, turbine arrangement, and channel dynamics on the performance of turbines in an array. We will answer the above questions and at the same time obtain some other important results.  Linear momentum actuator disk theory \citep{Nishino2012a} and the one-dimensional shallow-water equation \citep{Garrett2005} will be  used  to model an array in a channel. The main index, namely, power per turbine, will  be divided into two components: one is an indicator measuring the ability of turbines to acquire power from instantaneous flow and the other is the response of the channel to  turbines. We will also present results for the forces to which the turbines are subjected and environment influences, since these can compromise the power per turbine, with a consequent negative effect on the economics of building an array. 
 
For simplicity, we assume a constant cross section. The influence of channel constriction has been analyzed by \citet{Vennell2011} and  \citet{Smeaton2016}. These works mainly focus on channel-scale problems (the power potential of the whole channel). \citet{Vennell2011} gave some examples of the influence of constriction on power per turbine, and more work on this aspect is expected to be done in the future to obtain general results. As well as constriction, we also ignore the effects of  control of turbine blades. 
\citet{Vennell2014} showed that changing drag coefficients could store energy in the tidal channel system and at the same time produce more power. \citet{Vennell2016} systematically compared a smart patient strategy (allowing  changes in the tuning parameters) with a patient strategy (constant tuning parameters) and an impatient strategy (maximizing the possible power at every moment rather than in a tidal cycle), and demonstrated that the impatient strategy produces much less power than the two patient strategies. Arrays using the smart patient strategy perform significantly better than those using the patient strategy when it is necessary to meet conditions on the timing of energy demands. However, from a power perspective, these strategies give nearly the same results in practice if the maximum drag is constrained. As we are  concerned mainly with power in this study, we assume constant tuning parameters.  

\section{Mathematical model}
\subsection{Channel model}

To solve the problem of an array in a tidal channel,  we first model a channel connecting two open water areas (Fig. 1). Following \citet{Garrett2005}, the one-dimensional shallow-water momentum equation will be used here. The geometry of the channel (of length $L$, width $W$, and depth $H$) is assumed to be invariable along the streamwise direction. The deployment of turbines is modeled as an added drag in the momentum equation: 
\begin{equation}
\frac{\partial u}{\partial t}+u\frac{\partial u}{\partial x}=-g\frac{\partial \eta }{\partial x}-{{F}_{D}}-{{F}_{T}},
\end{equation}
where $u$ is the flow velocity in the channel and $\eta$ is the free-surface elevation. As the channel is considered to be dynamically short, the velocity and  cross-sectional area of the flow are constant here. The momentum losses $F_D$ and $F_T$ respectively represent the bottom drag and the turbine drag acting at the positions of the turbines. 
\begin{figure}
 \centerline{\includegraphics[width=12cm]{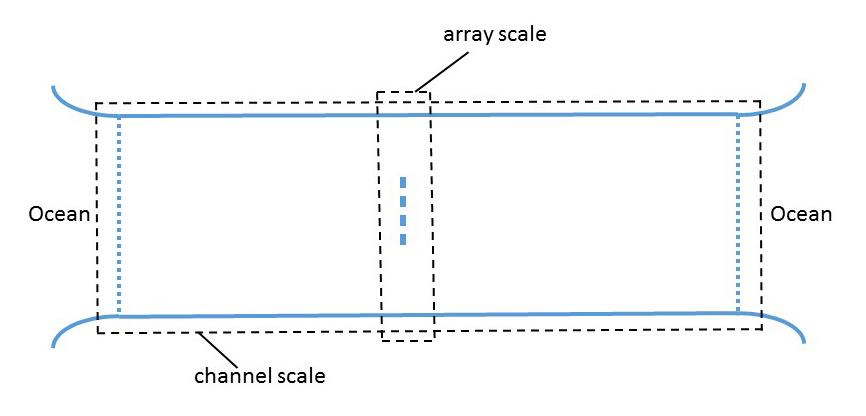}}
 \caption{Schematic of the simplified channel.}
\end{figure}

Ignoring  separation at the channel exit, the balance of forces in the tidal channel can be obtained by integrating the momentum equation along the streamwise direction:
\begin{equation}
\frac{\partial u}{\partial t}=g\left( \eta {{|}_{x=0}}-\eta {{|}_{x=L}} \right)/L-{{C}_{D}}\frac{u\left| u \right|}{H}-{{C}_{T}}\frac{u\left| u \right|}{L},
\end{equation}
where $C_D$ is the bottom drag coefficient and $C_T$  is the drag coefficient caused by the turbines. The first term represents the acceleration of the flow, and the second is the driving force resulting from the water-level difference. The last two terms are drag caused by bottom friction and the deployment of the turbines. 
\begin{table}
\caption{Parameters of  three hypothetical channels. \citep{Vennell2016}}
\begin{ruledtabular}
\begin{tabular}{lcccc}
&Small channel&Medium channel&Large channel\\
\hline
Length, Width, Depth & 4 km, 1.8 km, 10 m & 20 km, 9 km, 50 m & 60 km, 27 km, 150 m\\
Headloss amplitude & 0.56 m & 0.9 m & 2.2 m\\
 $C_D$ & 0.0025 & 0.0025 & 0.0025\\
 $\alpha =gA/\omega^{2}{L}^{2}$ & 17 & 1.1 &0.3\\ 
$\lambda_{D}=\alpha C_{D}L/H$ & 17 & 1.1 &0.3\\ 
\end{tabular}
\end{ruledtabular}
\end{table}

As before, we do not consider the influence of the turbines on open water areas (although the extraction of energy could cause changes in tidal amplitudes\citep{Karsten2013}). The effective length is the geometrical length of the channel. Assuming a sinusoidal head loss between the two sides of the channel, $\eta {{|}_{x=0}}-\eta {{|}_{x=L}}=A \sin \omega t~$  (the situation of compound tides has been analyzed by \citet{Adcock2014}), Eq. (2) can be nondimensionalized as 
\begin{equation}
\frac{\partial u '}{\partial t '}= \sin{t'}-(\lambda_D+\lambda_T)|u'|u',
\end{equation}
where $u'=u/u_\mathrm{max}$ is the nondimensional velocity ($u_\mathrm{max}$ is the velocity amplitude for a channel with no drag). $\lambda_D=\alpha C_D L/H$ and $\lambda_T=\alpha C_T $ are nondimensional drag coefficients, with $\alpha =gA/\omega^{2}L^{2}$. The time $t$ is nondimensionalized by $1/\omega$  and is therefore measured in radians. $\lambda_D$ and $\alpha$ are determined by environmental data. 
It should be noted that Eq. (3) is actually zero-dimensional yet it is still referred to as one-dimensional channel model following previous work (  \citet{Smeaton2016} and \citet{Vennell2016}) as it is given by one-dimensional shallow-water momentum equation. 
In this work, the background channel examples are the hypothetical small, medium, and large channels of \citet{Vennell2016}, as shown in Table I.

\subsection{Array model}

\begin{figure}
  \centerline{\includegraphics[width=12cm]{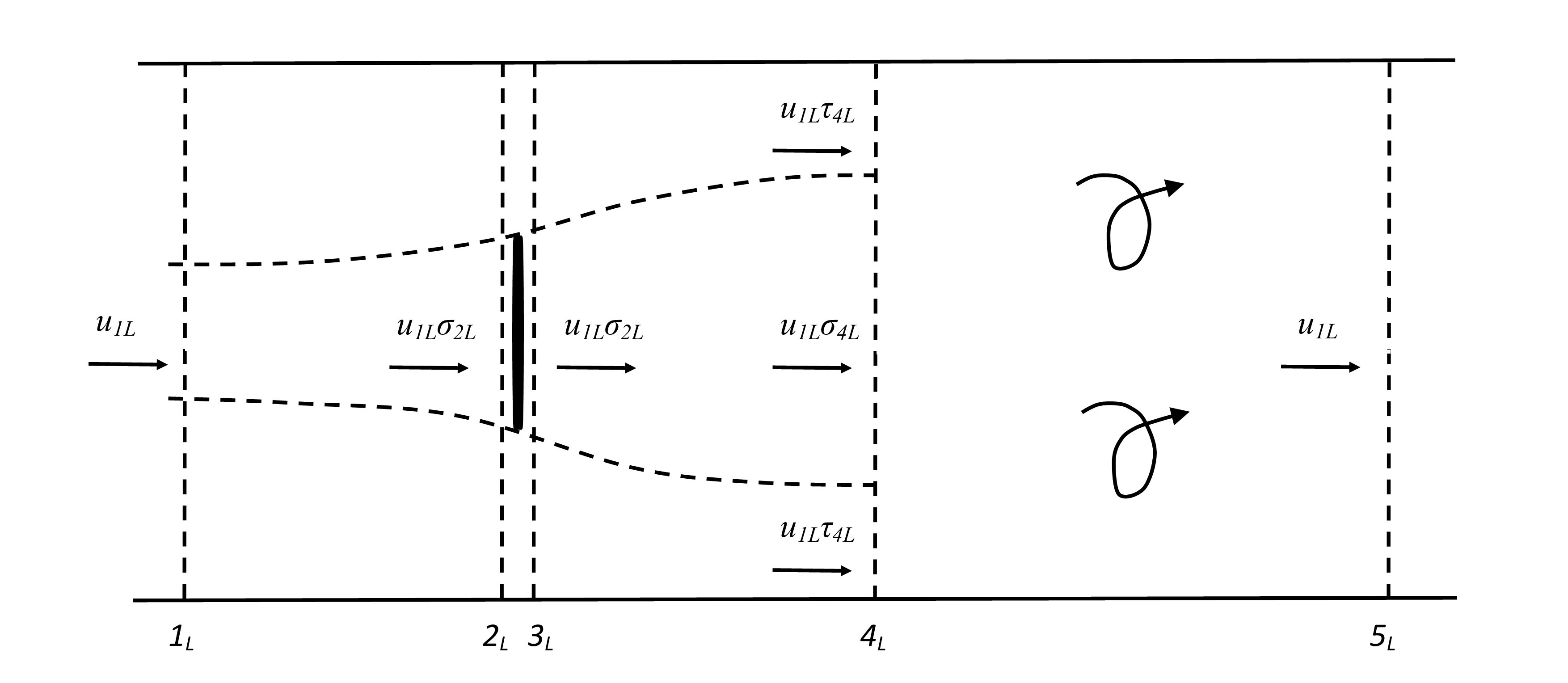}}
  \centerline{(a)}
  \centerline{\includegraphics[width=12cm]{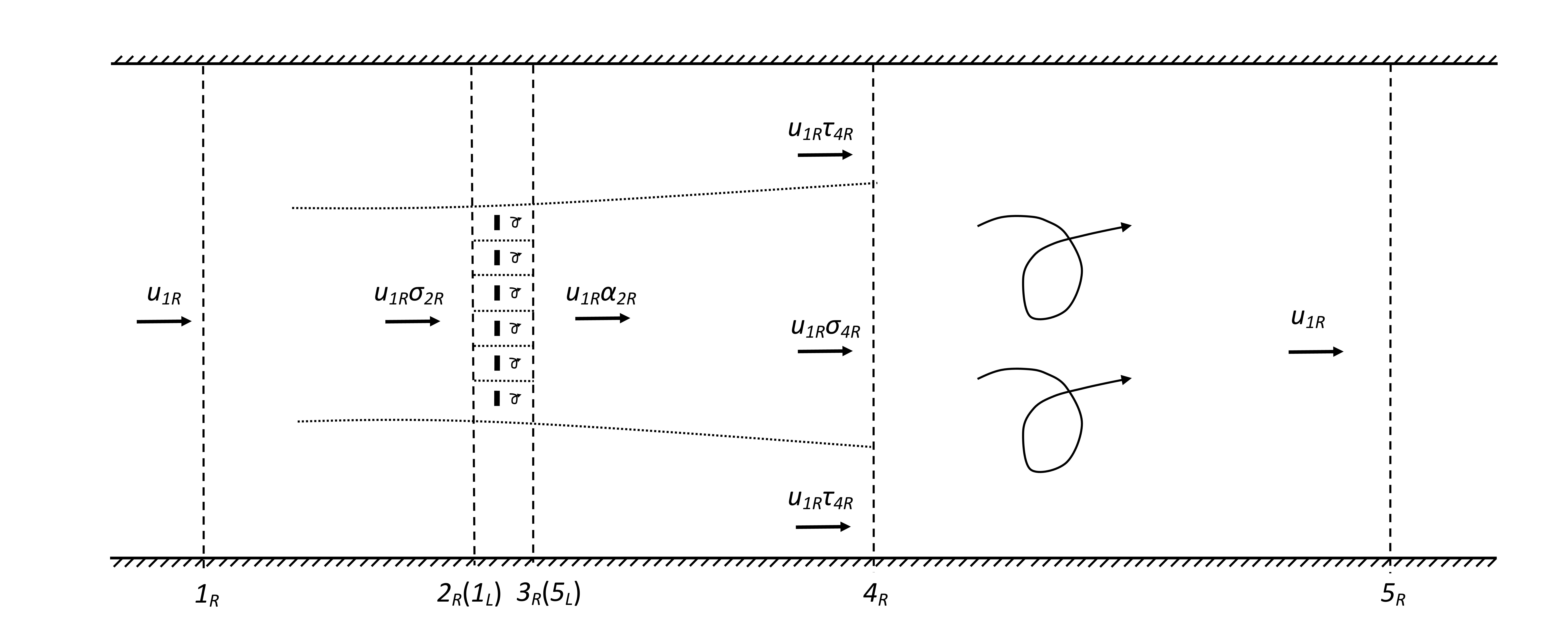}}
  \centerline{(b)}
  \caption{Schematics of the turbine-scale (a) and row-scale (b) actuator disk model for partial arrays. Neglecting the subscripts $L$, (a) can also be used to illustrate the dynamics of an array spanning the whole width of a channel.}
\end{figure}

Besides the channel model, we also need an array model that describes the dynamics within the array. Different rows in the array are assumed to be unaffected by the wakes of others. Thus, the single-row array model can be directly extended to model arrays with multiple rows of the same type. This assumption will be discussed in Sec. V. Based on the linear momentum actuator disc theory, the turbine is considered as a momentum sink. The flow is  divided into two components: the core flow through the turbines and the bypass flow. $\sigma$ and $\tau$ respectively denote the ratios of the core and bypass velocities to the upstream velocity. Usually, five stations are specified  as shown in Fig. 2, with 1 to 5 respectively representing the upstream station, the station immediately before the turbine, the station immediately after the turbine, the near-downstream station where the pressure equilibrates, and the far-downstream station. In the array model, we assume that the bottom drag is small compared with the turbine force. Thus, we neglect the effects of the bottom drag on the power coefficients and the effects of its variation between the core flow and bypass flow on the total drag. The influence of the bottom drag has been studied by \citet{Creed2017} and \citet{Gupta2017}, and we will also discuss it in the last section.   

First, we consider the dynamics of a whole-width row to show the effects of blockage. The global blockage, which controls the blockage effects, is defined as the ratio of the total turbine area to the channel cross-sectional area:   
\begin{equation}
B_G=\frac{n A_T}{WH},
\end{equation}
where $A_T$ is the turbine sweep area and $n$ is the number of turbines in one row. 

Following GC07, using the conservation of mass, momentum, and energy, the global force coefficient is 
\begin{equation}
{{C}_{FG}}=\frac{nF}{\frac{1}{2}\rho u_{1}^{2}n{{A}_{T}}}=\left( 1-{{\sigma }_{4}} \right)\frac{ \left( 1+{{\sigma }_{4}} \right)-2{{B}_{G}}{{\sigma }_{2}} }{{{\left( 1-{{B}_{G}}{{\sigma }_{2}}/{{\sigma }_{4}} \right)}^{2}}},
\end{equation}
where $F$ is the turbine force. The two nondimensional velocities $\sigma_2$ and $\sigma_4$ are  related by
\begin{equation}
{{\sigma }_{2}}=\frac{1+{{\sigma }_{4}}}{\left( 1+{{{B}}_{G}} \right)+\sqrt{{{\left( 1-{{{B}}_{G}} \right)}^{2}}+{{{B}}_{G}}{{\left( 1-1/{{\sigma }_{4}} \right)}^{2}}}}.
\end{equation}
The global power coefficient is
\begin{equation}
{{C}_{PG}}=\frac{n{{P}_{T}}}{\frac{1}{2}\rho u_{1}^{3}n{{A}_{T}}}={{C}_{FG}}{{\sigma }_{2}},
\end{equation}
where $P_T=F u_T$ is the power extracted by each turbine and $u_T$ is the velocity of the flow through the turbines.

The efficiency, which is defined as the ratio of the power extracted by the turbines to the total power removed in the flow caused by the turbines, is
\begin{equation}
\varepsilon_G=\sigma_2
\end{equation}

In practice, a row of turbines  can rarely be uniformly arranged across the whole width of the channel.  For rows partially blocking the channel, the flow becomes two-scale, and is represented by flows around the row [Fig. 2(b)] and around the turbine [Fig. 2(a)]. The basic dynamics of these two flow problems are the same. Thus, the GC07 model can be applied to both the local-scale and row-scale problems. Subscripts $L$ and $R$ are introduced to denote the local-scale and row-scale problems, respectively. 

The local blockage that controls the arrangement of the row is the ratio of the area of a single turbine  to the cross-sectional area of the local passage:
\begin{equation}
{{B}_{L}}=\frac{\frac{1}{4}\pi {{d}^{2}}}{\left( s+d \right)H},
\end{equation}
where $d$ is the turbine diameter and $s$ is the lateral inter-turbine spacing. 
Similarly, the row blockage is the frontal area of the row-scale fence normalized by the cross-sectional area of the channel: 
\begin{equation}
{{B}_{R}}=\frac{n\left( s+d \right)H}{WH}.
\end{equation}
The global blockage is the product of the local  and row blockages:  
\begin{equation}
{{B}_{G}}=\frac{n\frac{1}{4}\pi {{d}^{2}}}{WH}={{B}_{L}}{{B}_{R}}.
\end{equation}

For local-scale dynamics, we have
\begin{equation}
{{C}_{FL}}=\left( 1-{{\sigma }_{4L}} \right)\frac{ \left( 1+{{\sigma }_{4L}} \right)-2{{B}_{L}}{{\sigma }_{2L}}}{{{\left( 1-{{B}_{L}}{{\sigma }_{2L}}/{{\sigma }_{4L}} \right)}^{2}}},
\end{equation}
\begin{equation}
{{\sigma }_{2L}}=\frac{1+{{\sigma }_{4L}}}{\left( 1+{{B}_{L}} \right)+\sqrt{{{\left( 1-{{B}_{L}} \right)}^{2}}+{{B}_{L}}{{\left( 1-1/{{\sigma }_{4L}} \right)}^{2}}}},
\end{equation}
while for row-scale dynamics, we have
\begin{equation}
{{C}_{FR}}=\left( 1-{{\sigma }_{4R}} \right)\frac{ \left( 1+{{\sigma }_{4R}} \right)-2{{B}_{L}}{{\sigma }_{2R}} }{{{\left( 1-{{B}_{R}}{{\sigma }_{2R}}/{{\sigma }_{4R}} \right)}^{2}}},
\end{equation}
\begin{equation}
{{\sigma }_{2R}}=\frac{1+{{\sigma }_{4R}}}{\left( 1+{{B}_{R}} \right)+\sqrt{{{\left( 1-{{B}_{R}} \right)}^{2}}+{{B}_{R}}{{\left( 1-1/{{\sigma }_{4R}} \right)}^{2}}}}.
\end{equation}
These two problems are coupled by a continuity condition and a dynamical condition. The continuity condition is naturally encapsulated in the model:  
\begin{equation}
{{u}_{2R}}={{u}_{1L}},\qquad{{u}_{3R}}={{u}_{5L}}.
\end{equation}

Considering that the turbine thrust is the only force in the flow domain, the dynamical condition requires that the total forces at all scales must be the same: 
\begin{equation}
F_R=n F.
\end{equation}
Thus, the two force coefficients are related by
\begin{equation}
{{C}_{FR}}={{C}_{FL}}{{B}_{L}}\sigma _{2R}^{2}.
\end{equation}

The full equation system for the partial-array model consists of five equations (12)--(15) and (18) and six variables. When one variable is fixed, the full equation system can be solved. The global force coefficient, power coefficient, and efficiency are respectively 
\begin{equation}
{{C}_{FG}}=\frac{n{{F}_{L}}}{\frac{1}{2}\rho u_{1R}^{2}n\frac{1}{4}\pi {{d}^{2}}}={{C}_{FL}}\sigma _{2R}^{2},
\end{equation}
\begin{equation}
{{C}_{PG}}=\frac{n{{P}_{L}}}{\frac{1}{2}\rho u_{1R}^{3}n\frac{1}{4}\pi {{d}^{2}}}={{C}_{FL}}{{\sigma }_{2L}}\sigma _{2R}^{3},
\end{equation}
\begin{equation}
\varepsilon_G=\sigma_{2L} \sigma_{2R}.
\end{equation}

\subsection{Combined model}
According to the definition of the drag coefficient, the turbine drag coefficient $C_T$ in Eq. (2) can be determined by the array model: 
\begin{equation}
{{C}_{T}}=\tfrac{1}{2}{{N}_{R}}{{B}_{G}}{{C}_{FG}},
\end{equation}
where $N_R$ is the number of rows in the array.

The power per turbine can be deduced from two expressions:
\begin{equation}
{{P}_{T}}=\frac{1}{n{{N}_{R}}}\rho {{C}_{T}}A{{u}^{2}}\left| u \right|{{\varepsilon }_{G}}=\frac{1}{2n}\rho {{C}_{PG}}{{B}_{G}}A{{u}^{2}}\left| u \right|.
\end{equation}

The nondimensional power per turbine $\overline{P_T}'$ is calculated as the ratio of the average power of each turbine in a tidal cycle to the  kinematic flux of the blocking area before the turbine deployment: 
\begin{equation}
\overline{{{P}_{T}}}'=\frac{\mathop{\int }_{t}^{{2\pi }/{\omega }+t}{{P}_{T}}\,dt}{\mathop{\int }_{t}^{{2\pi }/{\omega }+t}\frac{1}{2}\rho {{\left| {{u}_{0}} \right|}^{3}}{{A}_{T}}\,dt},
\end{equation}
where $u_0$ is the undisturbed velocity before the turbine deployment. A factor $\omega /{2 \pi }$ has been omitted from the numerator and denominator. For convenience in the following discussion, we will  use the power per turbine directly to represent $\overline{P_T}'$. 

As we assume a constant tuning parameter, $C_{PG}$ is unchanged during a tidal cycle. So $\overline{P_T}'$ is found to be given by  
\begin{equation}
\overline{P_T}'=\gamma C_{PG},
\end{equation}
where $\gamma =\overline{|u'|^3}/\overline{|{u_0}'|^3}$ is the environment coefficient. For a given $C_T$, $\gamma$ can be obtained from Eq. (3). The nonlinear equation (3) was solved numerically  using a fourth-order Runge--Kutta  algorithm. The analytical solution in \citet{Vennell2010} was used as an initial condition to ensure convergence. 

Equation (25) demonstrates that the power per turbine $\overline{P_T}'$ is determined by two factors: the power coefficient $C_{PG}$ and the environment coefficient $\gamma$. $C_{PG}$ can be obtained from array models. It measures the ability of the turbines to acquire power from the instantaneous flow.  Change in $C_{PG}$ show the effects of blockage and turbine arrangement.  The environment coefficient $\gamma$ represents the influence of the reduced velocity on the power per turbine, which is caused by the interaction between the added force and the flow. $\gamma$ is always less than 1 and its variations  are  contrary to those of the drag coefficient $C_T$. When the channel dynamics is neglected, $\overline{P_T}'$ becomes $C_{PG}$. 
\section{Effects of channel dynamics and blockage}

We first consider an array spanning the whole channel width to investigate the influence of channel dynamics and blockage.  Figure 3 gives an example of an array ($B_G=0.3$) in the small channel. The predicted power is smaller than in the situation of constant velocity amplitudes. At the same time, because the environment coefficient $\gamma$ always decreases with  increasing axial induction factor $a=1-\varepsilon_G$, the optimal axial induction factor to give the maximum $\overline{P_T}'$ will also decrease. 
\begin{figure}
\includegraphics[width=12cm]{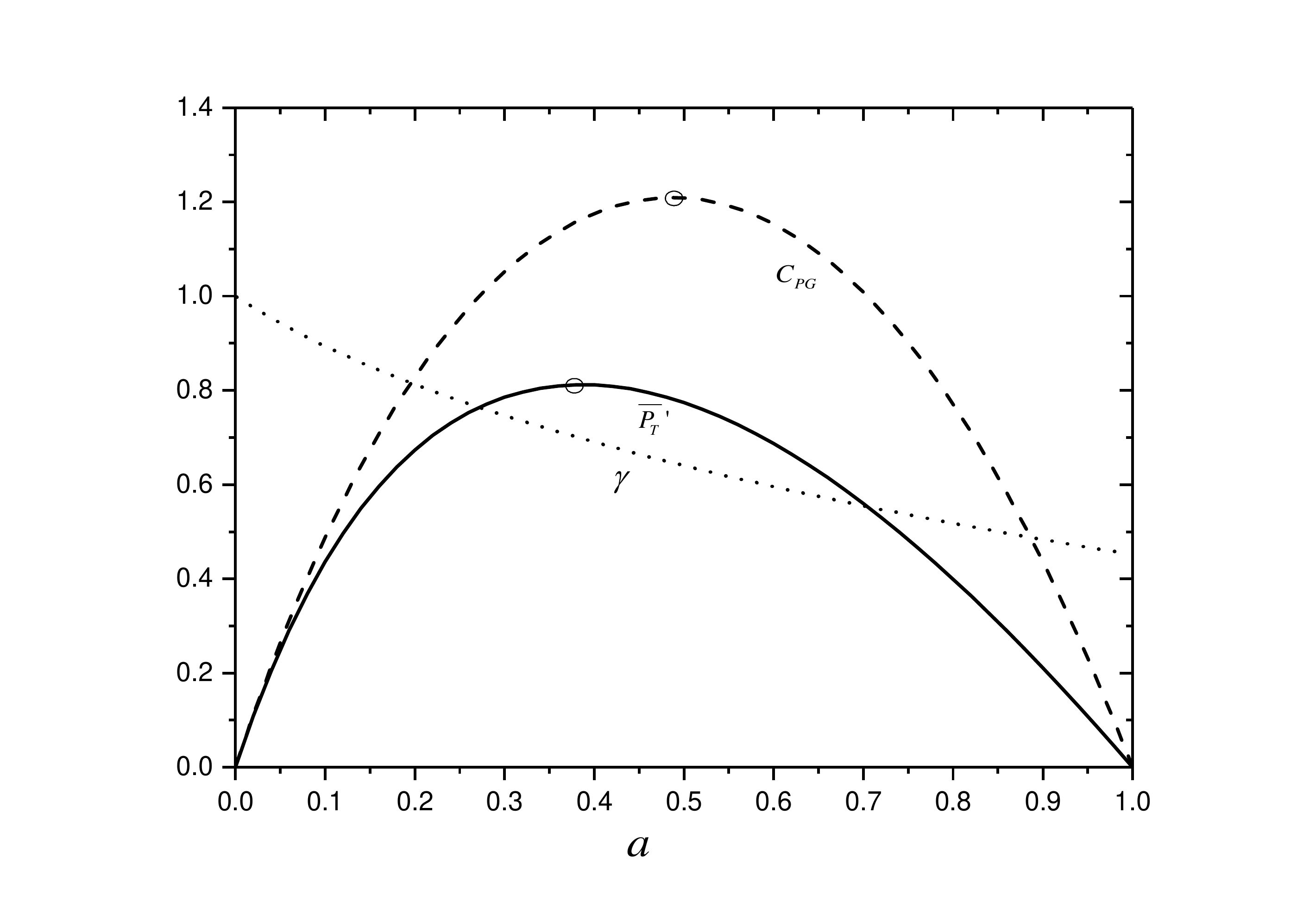}
\caption{The results of a single-row GC07 array ($B_G=0.3$) in the small channel versus the axial induction factor $a=1-\varepsilon_G$. The solid line shows the power per turbine $\overline{P_T}'$. The dashed and dotted lines show the two factors of $\overline{P_T}'$, namely, the power coefficient $C_{PG}$ and the environment coefficient $\gamma$, respectively.}
\end{figure}

Figure 3 shows the performance of the single-row array.  The numerical results show that among arrays with fixed global blockage, the single-row array has the largest $\overline{P_T}'$, although the total power may increase if more rows are added to the array.\citep{Vennell2010} This can also be derived from Eq. (25); for every possible $C_{PG}$ giving the maximum $\overline{P_T}'$, arrays with fewer rows always have smaller $C_T$ and thus larger $\gamma$ and $\overline{P_T}'$. When  the number of turbines is fixed, the situation becomes more complicated. \citet{Draper2014} have demonstrated that their conclusion regarding the sequence of arrangements is also valid when the influence of channel dynamics is taken into account. Figure 2 in \citet{Vennell2015} also shows that when the number of turbines is fixed, arrays with fewer rows have greater power potential. As we are mainly concerned here with the power per turbine, we will focus on single-row arrays.  

\begin{figure}
\includegraphics[width=12cm]{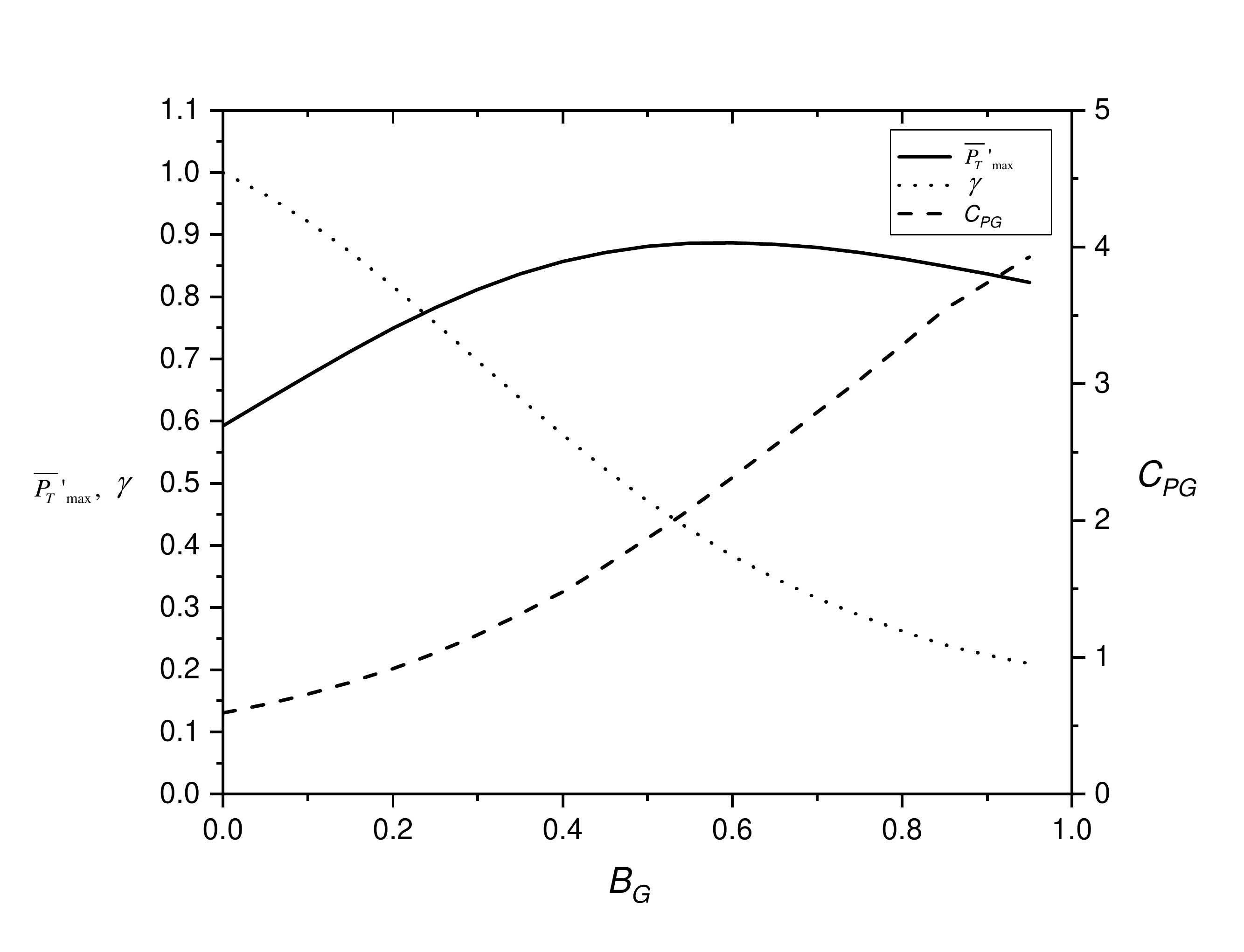}
\caption{Maximum power per turbine of a single-row GC07 array for a range of global blockages (solid line) and the corresponding $C_{PG}$ (dashed line) and $\gamma$ (dotted line).}
\end{figure}

The maximum $\overline{P_T}'$ together with the corresponding $C_{PG}$ and $\gamma$ for turbines in the small channel are shown in Fig. 4. For an array in a flow with constant velocity amplitude,  $\overline{P_T}'_\mathrm{max}$ (actually $C_{PG\mathrm{max}}$) will increase monotonically with the global blockage. When we consider the influence of $\gamma$,  there is an obvious maximum value ($B_G \approx 0.58$) in Fig. 4. Adding turbines to a cross section will initially enhance, then reduce, the performance of turbines (from a power perspective). Yet, for certain large channels, adding turbines will always increase  $\overline{P_T}'_\mathrm{max}$ (which then reaches its maximum value at $B_G=1$). There are also situations, for example the small channel in \citet{Vennell2011} (with $\lambda_D=4.6$ and $\alpha=18$),  where the addition of turbines will monotonically decrease  $\overline{P_T}'_\mathrm{max}$ (which then reaches its maximum value at $B_G=0$). 
\begin{figure}
\includegraphics[width=12cm]{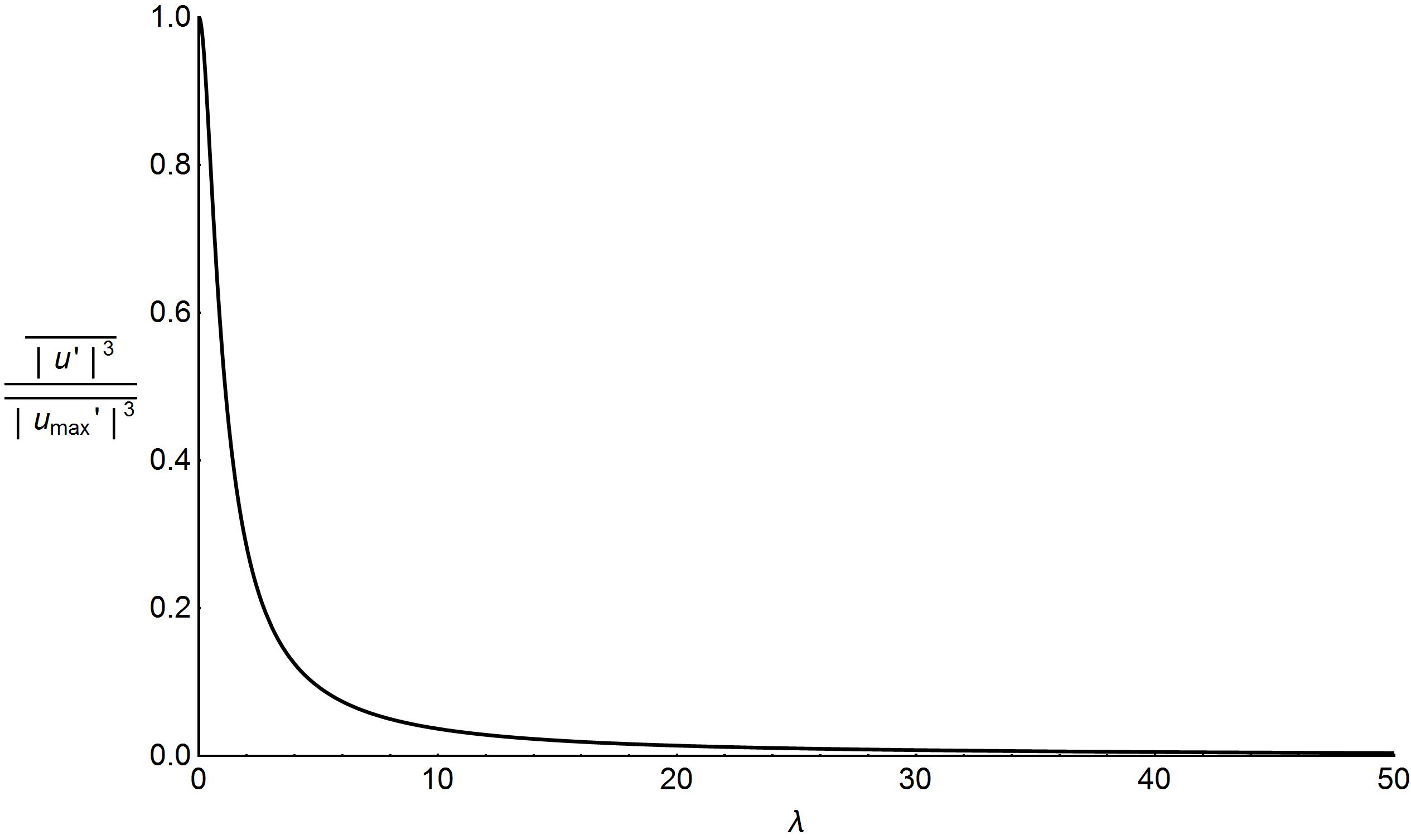}
\caption{Variation of the cube of the nondimensional velocity $\overline{|u'|^3}$ relative to the nondimensional drag coefficient $\lambda$. $\overline{|u'|^3}$ is divided by $\overline{|u'_\mathrm{max}|^3}=\int_0^{2\pi} \sin{t}\,dt/{2\pi}$ to ensure that the value of the curve is 1 at $\lambda=0$. In general, the slope of the curve increases with  increasing $\lambda$.}
\end{figure}
\begin{figure}
\includegraphics[width=14cm]{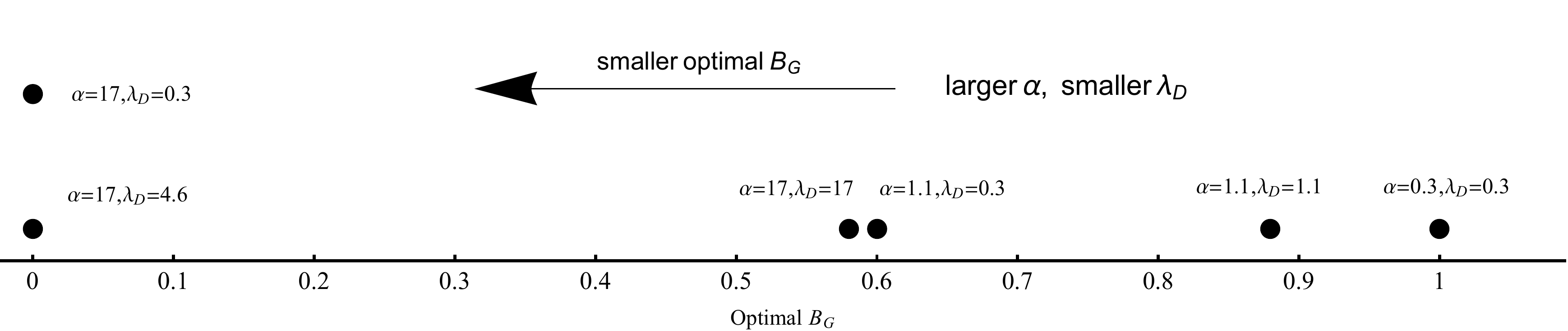}
\caption{The optimal global blockages (black solid points) to attain the extreme value of $\overline{P_T}'_\mathrm{max}$ for single-row arrays in channels with different $\alpha$ and $\lambda_D$. Arrays in channels with large $\alpha$ and small $\lambda_D$ have relatively small optimal $B_G$. }
\end{figure}

As we discussed earlier, optimal tuning requires a relatively large $C_{PG}$ while at the same time preventing $\gamma$ from becoming too small. 
In general, when $C_T$ is fixed, arrays with smaller $B_G$ have larger $C_{PG}$. Thus, when $B_G$ increases, the only way to obtain higher $\overline{P_T}'_\mathrm{max}$ is by increasing $C_{PG}$ ($C_T$) from its original optimal value. 
For small channels with large $\alpha$,  increasing $C_T$ will cause a large change in $\lambda=\lambda_D+\lambda_T$ and thus a serious reduction in velocity. At the same time, as the slope of  $\overline{|u'|^3}$  increases sharply with  increasing $\lambda$, when $\lambda_D$ is small, an increase in $\lambda$ will necessarily decrease the value of $\overline{|u'|^3}$ to a great ratio (Fig. 5). Thus, the different results of adding turbines are related to $\alpha$ and $\lambda_D$. These two parameters determine the relation between the turbine drag coefficient $C_T$ and the environment coefficient $\gamma$: the values of $\gamma$ for  arrays in channels with smaller $\lambda_D$ and larger $\alpha$ are more sensitive to  changes in  $C_T$. 

It is not easy to determine the quantitative effects of $\alpha$ and $\lambda_D$ on variations in $\overline{P_T}'_\mathrm{max}$, since both the array model and the channel model are nonlinear. However, from our analysis so far, we can clearly see that arrays in channels with smaller $\lambda_D$ and larger $\alpha$ will have a smaller optimal $B_G$ values. Physically, smaller $\lambda_D$ and larger $\alpha$ means that the importance of the turbine drag is greater. Some numerical results are shown in Fig. 6, exhibiting the same trend as in the above analyses. For example, compared with the situation with $\alpha=1.1$ and $\lambda_D=0.3$, arrays in a channel with $\alpha=17$ and $\lambda_D=0.3$ have much smaller optimal $B_G$ values, while arrays in a channel with  $\alpha=1.1$ and $\lambda_D=1.1$ have larger optimal $B_G$ values. Large arrays may include several rows. In such circumstances, $\alpha$ will increase to $N_R \alpha$ for an array with $N_R$ rows, and the optimal global blockage will accordingly decrease. 

\section{Benefits of an appropriate arrangement}
\begin{figure}
\includegraphics[width=12cm]{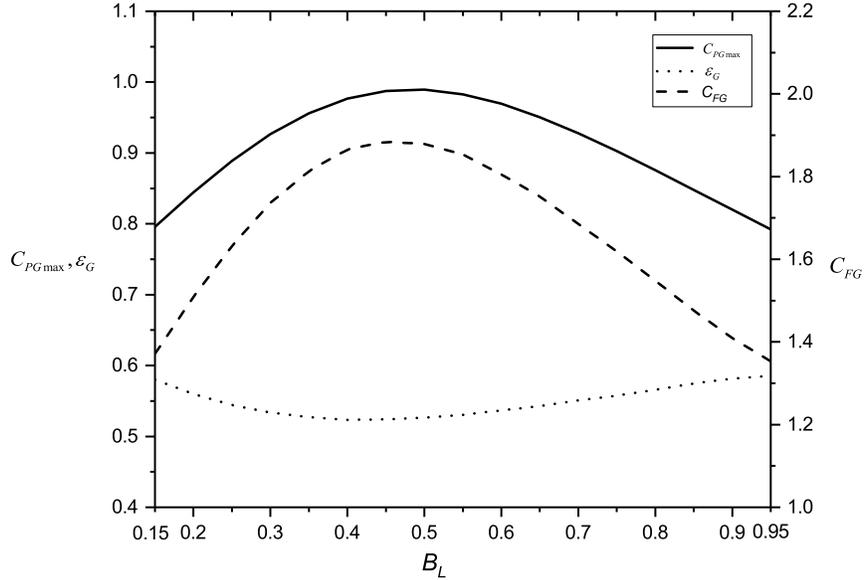}
\caption{Effects of local blockage on the maximum power coefficient (solid line) of an array with $B_G=0.12$ and constant velocity amplitudes, together with  the corresponding force coefficient (dashed line) and efficiency (dotted line).}
\end{figure}
We have demonstrated that the value and trend of variation of $\overline{P_T}'_\mathrm{max}$ depend on channel properties ($\lambda_D$ and $\alpha$) and blockage by turbines ($B_G$). According to NW12, under the assumption of constant flow, we can also benefit by arranging the turbines in an appropriate manner. 
\subsection{Qualitative analysis}
We first consider the question of whether the optimal arrangement of the array with constant flow is still appropriate when the dynamics of the channel are taken into account. This question can be solved by analyzing the parameters in the array model. 

Figure 7 shows an example of the changes in $C_{PG\mathrm{max}}$ as  the local blockage is altered. When $B_G$ is fixed as 0.12, the optimal arrangement corresponds to $B_L=0.48$. However, an increase in  maximum power is accompanied by an increase in force and a decrease in efficiency. 
Thus, it is hard to judge the performance of arrays with different arrangements just by comparing $C_{PG\mathrm{max}}$. 

\begin{figure}
  \centering
  \subfigure[]{
    \label{fig:subfig:a} 
    \includegraphics[width=7cm]{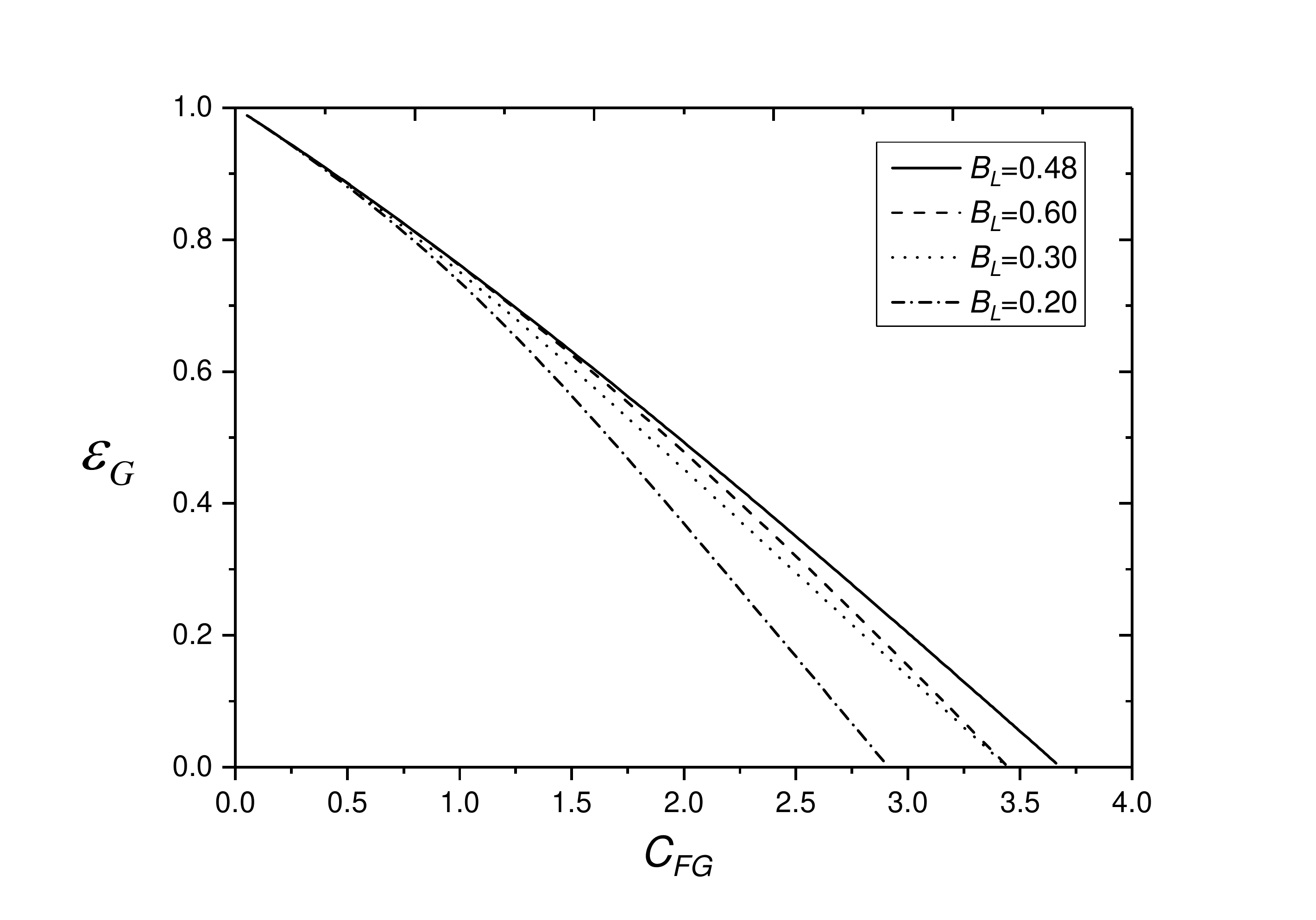}}
  \hspace{1cm}
  \subfigure[]{
    \label{fig:subfig:b} 
    \includegraphics[width=7cm]{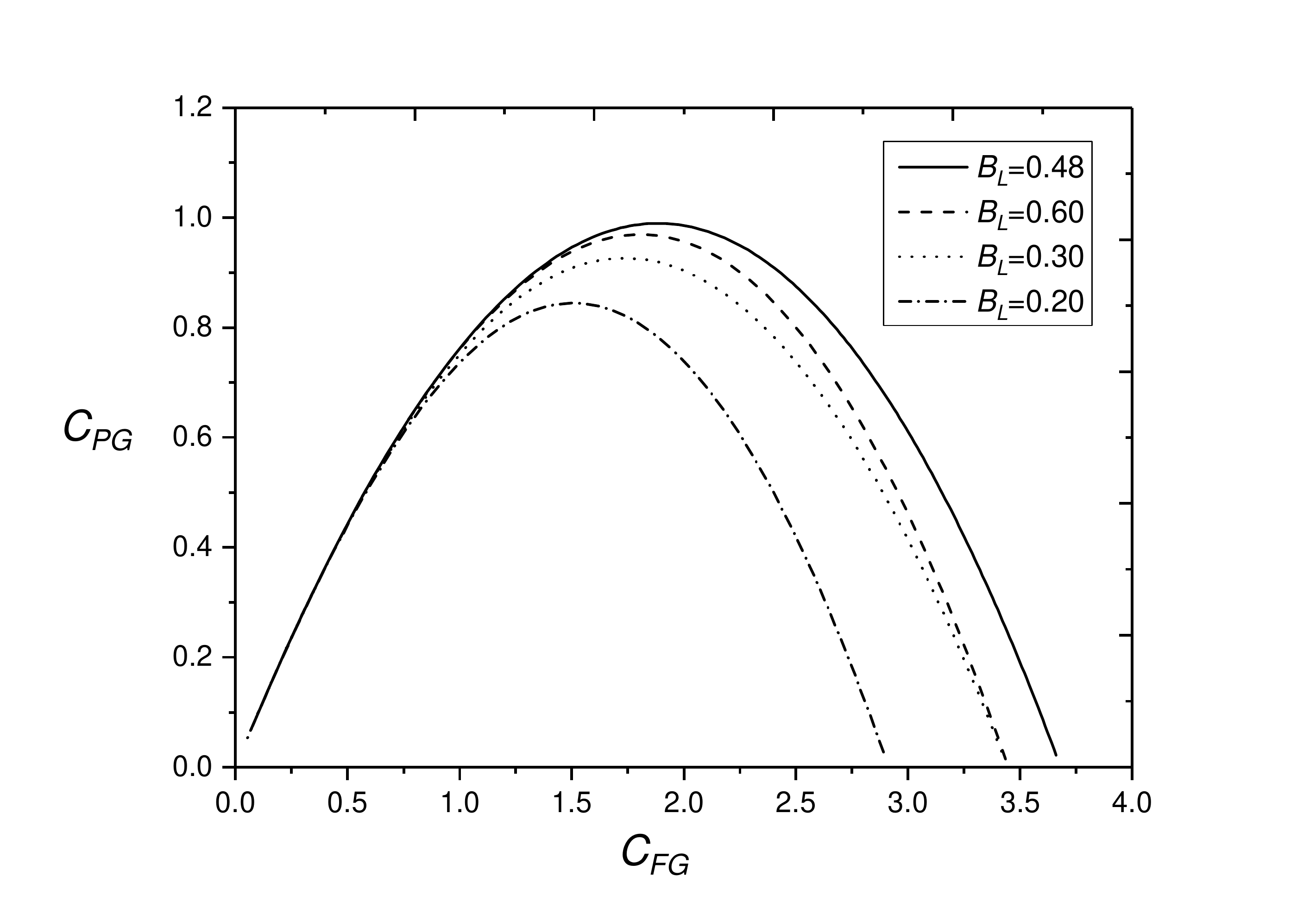}}
\caption{Efficiency $\varepsilon_G$ (a) and power coefficient $C_{PG}$ (b) as  functions of the global force coefficient $C_{FG}$ for arrays with $B_L=0.2$, 0.3, 0.48, and 0.6 when $B_G$ is fixed as 0.12.}
\end{figure}
Although we cannot determine the exact value of the power per turbine without a time series calculation, some inferences can still be made by analyzing certain variables. According to Eq. (3), the nondimensional drag coefficient $\lambda$ determines the velocity of the flow in the channel. Thus, for arrays with the same number of rows and global blockage, $\gamma$ is determined by $C_{FG}$ according to Eqs. (22) and (3). Following Eq. (25), the power per turbine can be expressed as a function of two variables $C_{FG}$  and $C_{PG}$, or $C_{FG}$ and $\varepsilon_G$:
\begin{equation}
\overline{{{P}_{T}}}=\mathcal{F}\left( {{C}_{FG}},{{C}_{PG}} \right)   \qquad \mathrm{or} \qquad
\overline{{{P}_{T}}}=\mathcal{F}\left( {{C}_{FG}},{{\varepsilon }_{G}} \right)
\end{equation}

Figure 8(a) shows the variation of efficiency $\varepsilon_G$ with the force coefficient $C_{FG}$ for arrays with $B_L=0.2$, 0.3, 0.48, and 0.6 when $B_G$ is fixed as 0.12. For any $C_{FG}$, the optimal arrangement $B_L=0.48$ has the largest $\varepsilon_G$. For other arrangements, such as $B_L=0.3$, the value of $C_{FG}$ that gives the maximum $\overline{P_T}'$ is uncertain. However, for all possible values of $C_{FG}$ that may correspond to $\overline{P_T}'_\mathrm{max}$, the array with $B_L=0.48$ always has a higher $\varepsilon_G$, leading to a  higher $\overline{P_T}'$. Thus, it can be deduced that the maximum power per turbine $\overline{P_T}'_\mathrm{max}$ of the array with the optimal arrangement will definitely be larger than that of others, even when we consider channel dynamics.  

Besides power, the force per turbine also plays an important role in the design of tidal arrays, especially from the point of view of economics: a large force may lead to a high cost. 
Furthermore, a comprehensive assessment of a project should include the influence of  turbines on tidal channels. These two aspects can be more clearly seen in Fig. 8(b). For a certain $C_{PG}$, the array with the optimal arrangement has the smallest $C_{FG}$, and thus the largest $\gamma$. As the nondimensional force $\lambda \overline{u{{'}^{2}}}$ exhibits the same trend as the nondimensional drag coefficient $\lambda$ and the opposite trend to the velocity, adding a force coefficient will increase the force, despite reducing the velocity. This means that the array with the optimal arrangement is able to produce more energy than arrays with other arrangements, while at the same time being subjected to lower forces and with less velocity reduction. However, as we do not know the $C_{FG}$ value giving  the maximum $\overline{P_T}'$,  it is hard to judge whether the velocity for the optimal arrangement corresponding to  maximum power is larger than the velocity for other arrangements. It should be noted that the previous statement that $\overline{P_T}'_\mathrm{max}$ of the array with the optimal arrangement is the largest can also be obtained from the curve of $C_{PG}$ versus $C_{FG}$.   

In summary, by merely using the array model and analyzing the variables from the perspective of the whole channel, we can obtain some qualitative conclusions. First, arrays with optimal arrangements have the largest power potential. Second, they have higher power--force efficiency and power--environment efficiency than others. However, uncertainties remain regarding the force and flow velocity corresponding to $\overline{P_T}'_\mathrm{max}$. 

\subsection{Quantitative analysis}
\begin{figure}
\includegraphics[width=12cm]{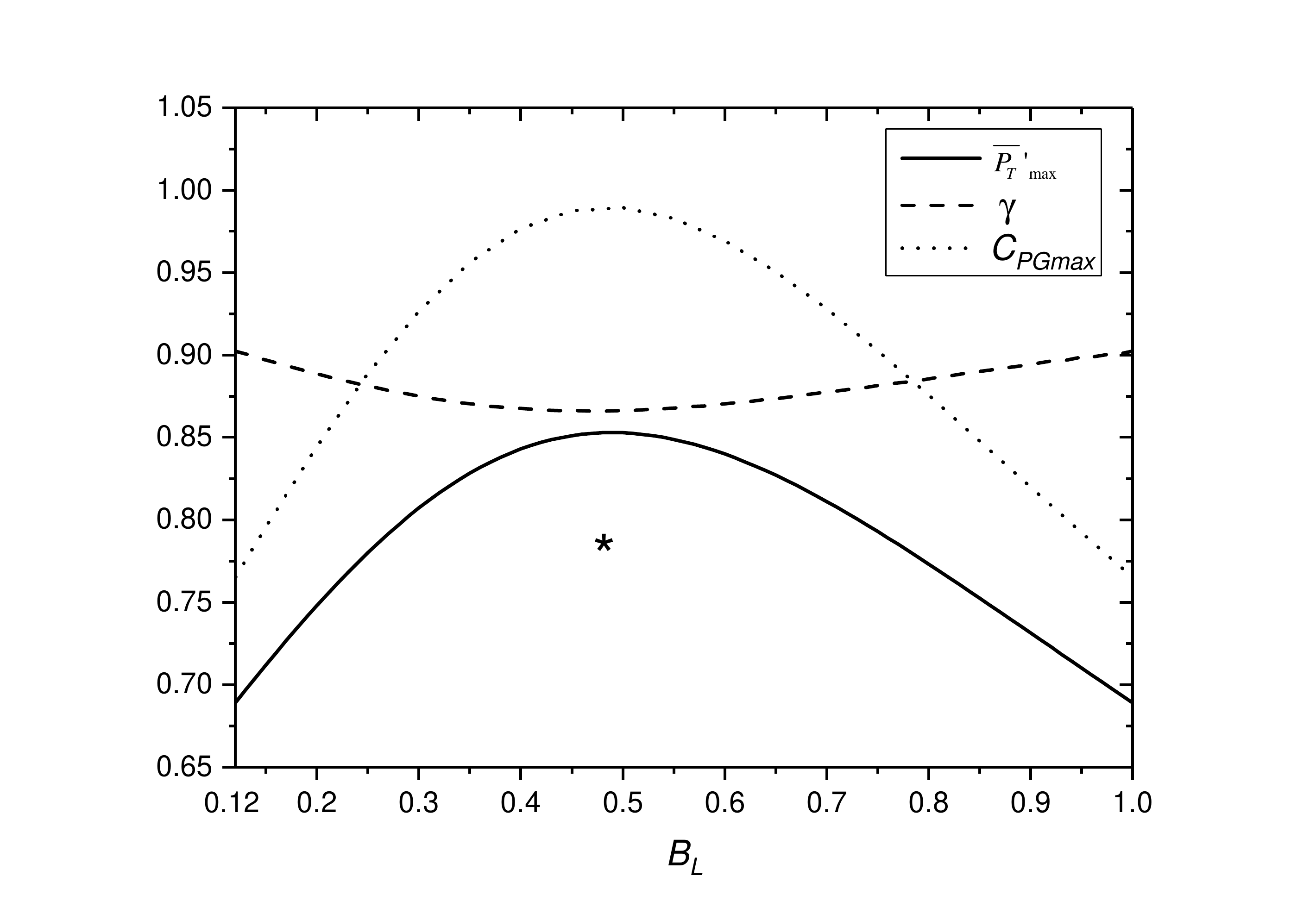}
\caption{Effects of local blockage on $\overline{P_T}'_\mathrm{max}$ of an array with $B_G=0.12$ in the small channel, together with the corresponding $\gamma$ and maximum $C_{PG}$. The star shows $\overline{P_T}'_\mathrm{max}$ for the array with optimal arrangement when $\gamma$ is restricted to be no less than 0.902.}
\end{figure}
If we want to know the extent to which an array can benefit from optimal blocking of the channel, as well as  the influence of the array with optimal tuning on the whole channel, we need to incorporate the partial array model into the channel dynamical model. Figure 9 shows the changes in $\overline{P_T}'_\mathrm{max}$ and the corresponding $\gamma$ versus the local blockage for an array  in the small channel. Corresponding to the previous analysis (Fig. 8), we still take the array with $B_G=0.12$ as an example. For comparison, $C_{PG\mathrm{max}}$ for constant velocity amplitudes is also shown is this figure. The left boundary $B_L=0.12$ and right boundary $B_L=1$ represent the points where the partial array reduces to the GC07 whole-width array. It is apparent that occupying the whole width is not a reasonable arrangement for an array. $\overline{P_T}'_\mathrm{max}$ reaches its  maximum value at $B_L=0.48$, which is the same as in the scenarios with constant velocity amplitudes, validating our previous qualitative variable analysis. It can also be seen that the trend of $\gamma$ versus local blockage is opposite to that of $\overline{P_T}'_\mathrm{max}$. Hence, when the full power potential is realized, arrays with optimal arrangements are subjected to larger forces and have greater influences on channels than arrays with other arrangements. 

In practice, the turbine force and velocity reduction may be restricted. Consistent with the qualitative analysis, the numerical results demonstrate that for the same force and the same reduced velocity ($\gamma$), arrays with optimal arrangements have the highest $\overline{P_T}'_\mathrm{max}$. For example, if $\gamma$ is restricted to be no less than 0.902, which is the $\gamma$ of $\overline{P_T}'_\mathrm{max}$ for a GC07 one-scale array, $\overline{P_T}'_\mathrm{max}$ for $B_L=0.48$ is 0.782. This value is still much larger than the one-scale result 0.689 at $B_L=0.12$ or $B_L=1$. 


\begin{figure}
\includegraphics[width=12cm]{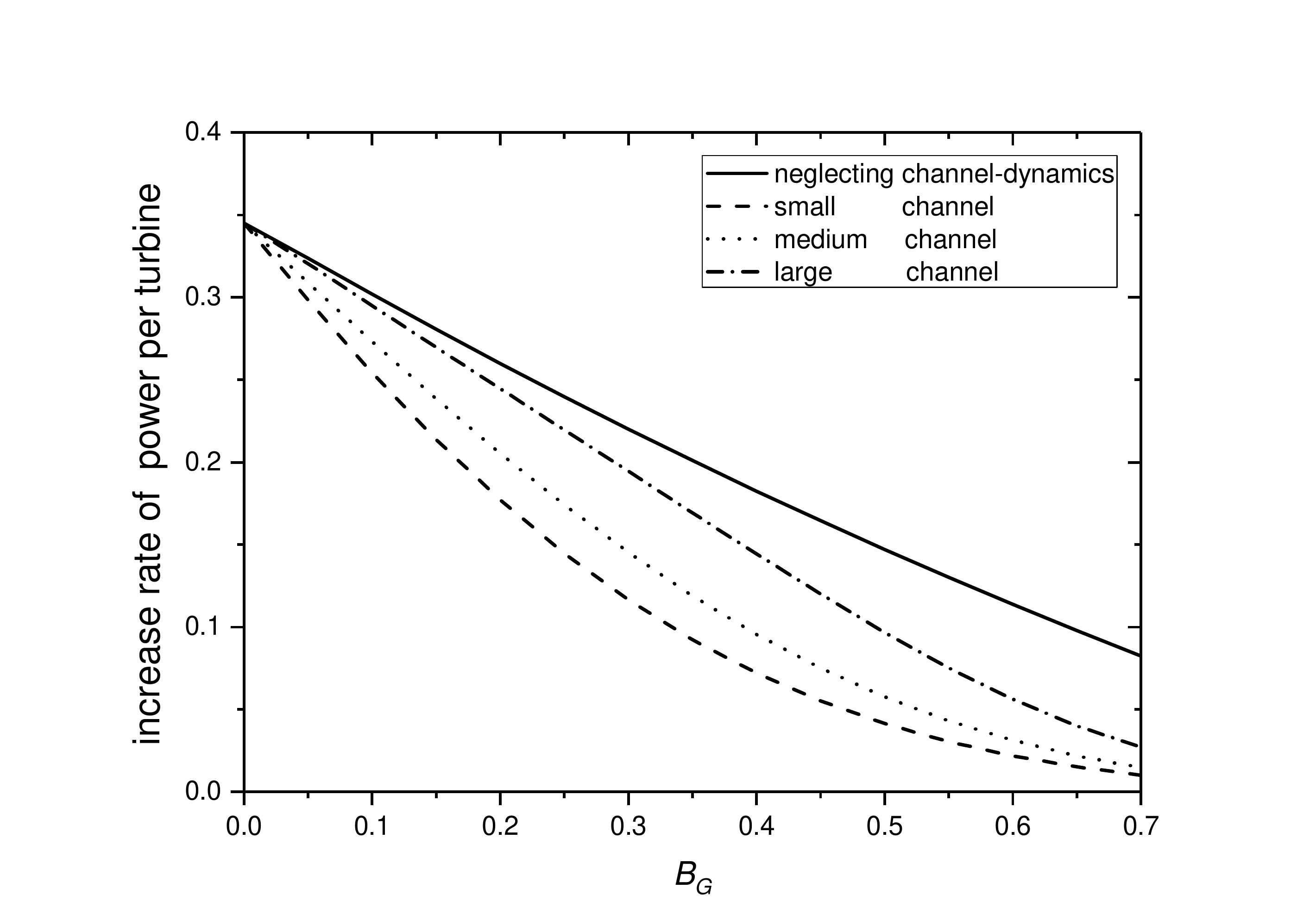}
\caption{The rate of increase  of $\overline{P_T}'_\mathrm{max}$ from an array spanning the whole channel width to an optimal partial array for a range of global blockages in four cases: the small, medium, and large channels and the channel with constant velocity amplitudes.}
\end{figure}
\begin{figure}
\includegraphics[width=12cm]{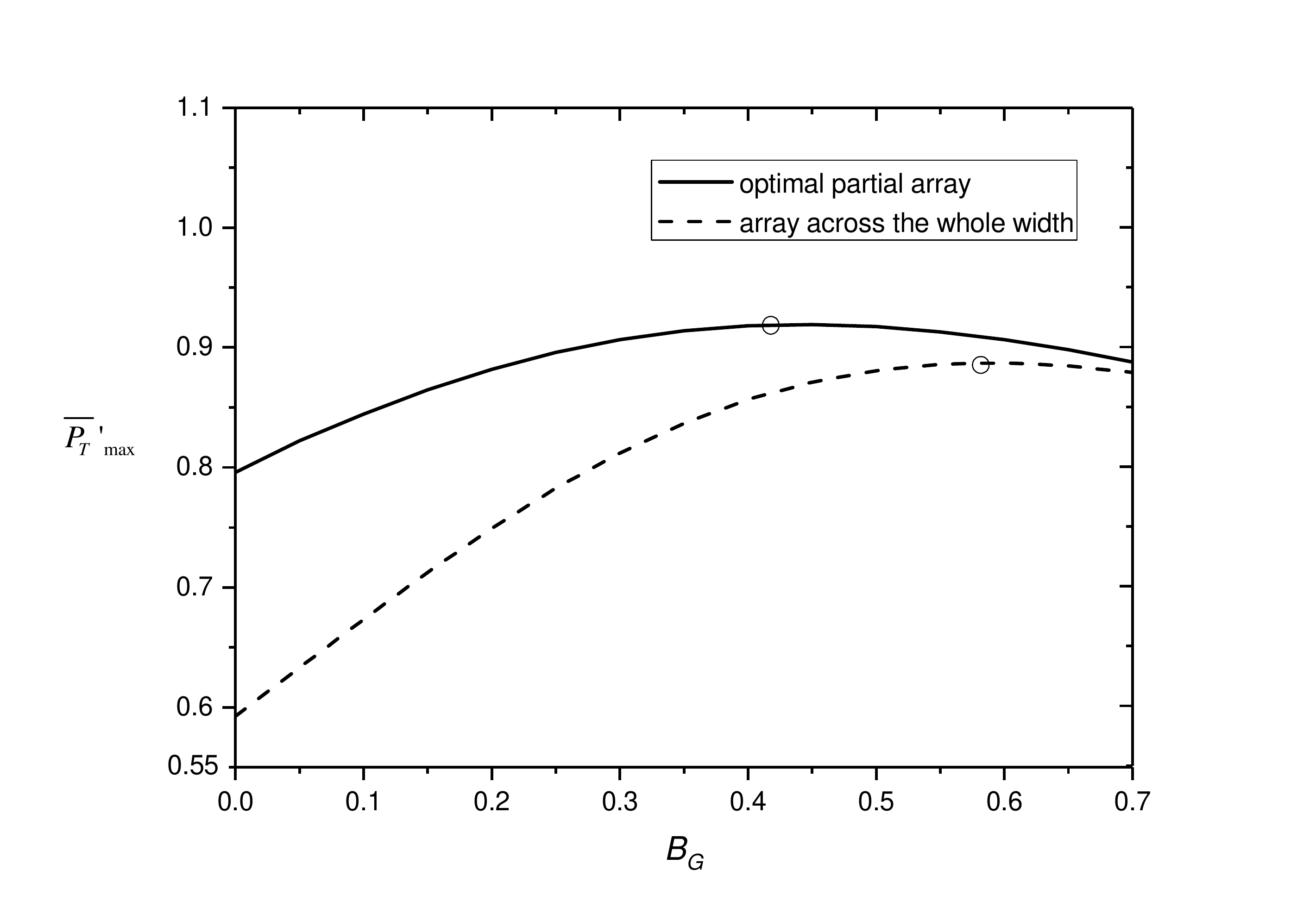}
\caption{$\overline{P_T}'_\mathrm{max}$ of arrays spanning the whole channel width and optimal partial arrays for a range of $B_G$.}
\end{figure}
For other global blockages in different channels, the numerical results confirm that arrays with optimal arrangements in situations with constant velocity amplitudes still have the maximum $\overline{P_T}'_\mathrm{max}$. It should be noted that the rate of increase  of $\overline{P_T}'_\mathrm{max}$ from arrays across the whole width of the channel to optimal partial arrays will decrease when the channel dynamics are taken into account (Fig. 10). Here we only show the results for $B_G<0.7$, because an extremely high blockage is not practical.  The small channel has the smallest growth rate. This sequence could also be attributed to the different sensitivities of the environment coefficient $\gamma$ and the drag coefficient $C_T$. However, for arrays with a moderate blockage, there are still noticeable benefits from an optimal arrangement, even in small channels.  For example, when $B_G$ is fixed as 0.3, a relatively high value, theoretically we can still get a $12\%$ increase in $\overline{P_T}'_\mathrm{max}$ in the small channel.

Despite the difference in values, the rate of increase will always decrease  with $B_G$, regardless of the channel parameters.  This is why the global blockage necessary for a partial array to achieve the extreme value of $\overline{P_T}'_\mathrm{max}$ will decrease compared with that for a one-scale array across the whole width. Figure 11 illustrates such a case. The array spanning the whole width of the small channel has its maximum  $\overline{P_T}'_\mathrm{max}$ at $B_G=0.58$, while for the partial array with the optimal arrangement, $\overline{P_T}'_\mathrm{max}$ will attain its maximum value at $B_G=0.43$ when a smaller number of turbines are deployed. 

\section{Conclusion and Discussion}

The performance of tidal turbine arrays can be judged from several perspectives. Previous work taking account of channel dynamics  focused mainly on the power potential of whole channels. For the commercial exploitation of tidal energy, the performance of tidal turbines is pivotal. In view of this, the power per turbine has been the main concern in this paper, since  it affects the performance and economics of tidal turbines. 

Previous studies of  array models have shown that the density and arrangement of turbines will influence their performance. The turbine density has already been incorporated  into the channel-dynamics model. However, the influence of the density on the trend of power per turbine needs more attention.
More importantly, to the best of our  knowledge, a gap exists between those studies that  stress different arrangements of arrays and those investigating the stress dynamics of channels. Thus, in this work, the influence of blockage, channel dynamics, and turbine arrangement have been simultaneously investigated. We have divided the power per turbine  $\overline{P_T}'$ into two components: an indicator measuring the ability of turbines to acquire power from the instantaneous flow (the power coefficient $C_{PG}$) and the response of the channel to turbines (the environment coefficient $\gamma$). In this way,  the influence of channel dynamics, together with the trade-off between the ability to acquiring power on the one hand and the reduced velocity on the other, is  obvious. 

The results show that the influence of channel dynamics will decrease the predicted power and the optimal induction factor compared with the case of constant velocity amplitudes. These decreases are not the same for  different channels. Thus, an interesting phenomenon appears. Depending on the channel conditions, when we gradually increase the global blockage from zero to one, the power per turbine will  monotonically increase or decrease or will reach a maximum at a certain $B_G$ (neglecting the channel dynamics, the trend is a monotonic increase). Specifically, arrays in channels with smaller $\lambda_D$ and larger $\alpha$ will require a smaller global blockage in orderto achieve the extreme value of $\overline{P_T}'_\mathrm{max}$, since the environment coefficient is more sensitive to the drag coefficient in such channels.

We also find that arrays can benefit greatly from an appropriate arrangement of turbines rather than spanning the whole channel width, even when  the channel dynamics are taken into account. 
The optimal arrangement for each turbine density derived from the two-scale model under constant situation still has the largest power potential. Furthermore, optimal partial arrays have better power--force and power--environment efficiency. However, it should be noted that the rate of increase in the maximum power per turbine from GC07 one-scale array to optimal partial array will decrease when the response of the channel flow to the turbines is taken into account. Finally, it has been shown that besides increasing $\alpha$ and decreasing $\lambda_D$, an optimal arrangement of the turbines in an array can also reduce the global blockage that gives the maximum power per turbine. Neglecting the influence of the turbine arrangement would lead to a departure from the optimal turbine density and a lower predicted power.

It is worth noting that the model adopted here provides a relatively simple description of  arrays in tidal channels. In practice, it is hard to predict the power potential using theoretical models. For example, real tidal channels are far more complicated than the hypothetical channels used here. The velocity also changes along with the depth and width. \citet{Draper2016} demonstrated that the power of a turbine will be influenced by the velocity profile of the channel and the relative positions of the turbines. The actuator disk model used here appears to provide an upper limit on the extractable power. In practice, tidal turbines consist of blades and cannot  produce the  power predicted by this model. Further numerical simulations and experiments are needed to estimate the performance of  real arrays. However, the aim of this work was not to provide precise results but rather to give insights into the dynamics of arrays in tidal channels and provide guidance for the design of efficient arrays.  

An important assumption in this work is that the bottom drag can be neglected within the area of an array. When turbines are arranged across the whole width of a channel, it is reasonable to neglect the bottom drag, since the device-scale flow area is negligible compared with that of the whole channel. For partial arrays occupying a large portion of the channel width, the array-scale flow area may be large enough to give rise to changes in the bottom drag. \citet{Gupta2017} and \citet{Creed2017} have developed models to investigate this. The influence of the bottom drag on turbine performance comes mainly from two aspects. First, it can directly change the dynamics within the array, increasing the power coefficient. Furthermore, the flow will also be affected by variations in drag. Including the effects of changes in the bottom drag  is a topic for future investigation. For partial arrays, another issue is the direct interaction between the rows of the array. We have assumed that each row will not be affected by the wakes of other rows, an assumption that requires relatively large distances between rows. Because of the expansion of the wake areas, the back rows of an array may be located in the wakes of the front rows, thus leading to a decrease in the power of their turbines, as shown in \citet{Draper2014}. 

It is also worth noting that the partial array model assumes steady flow while the channel model is time-dependent. Yet the time scale of the tide is far larger than of the array scale mixing. So it is reasonable to combine the partial array model and channel model considering the theoretical model itself is an approximation to physics. For partial arrays in a tidal channel, the partial array model is more suitable than GC07 one-scale model.  

\begin{acknowledgments}
The authors wish to acknowledge the support of the 1000 Talent Plan
\end{acknowledgments}

\nocite{*}
\bibliography{aipsamp}
\end{document}